\documentclass[journal,comsoc]{IEEEtran}

\usepackage{color}
\usepackage{amsmath,amsfonts}
\usepackage{algorithmic}
\usepackage{array}
\usepackage[caption=false,font=normalsize,labelfont=sf,textfont=sf]{subfig}
\usepackage{textcomp}
\usepackage{stfloats}
\usepackage{url}
\usepackage{verbatim}
\usepackage{graphicx}
\usepackage{tabularx}
\usepackage{makecell}
\usepackage{multirow}
\usepackage{threeparttable}
\usepackage{float}
\def\BibTeX{{\rm B\kern-.05em{\sc i\kern-.025em b}\kern-.08em
    T\kern-.1667em\lower.7ex\hbox{E}\kern-.125emX}}
\usepackage{balance}

\usepackage[absolute]{textpos}

\begin{document}

\begin{textblock}{13}(1.5,0.35)
	\noindent S. Sun, R. Li, C. Han, X. Liu, L. Xue, and M. Tao, ``How to differentiate between near field and far field: Revisiting the Rayleigh distance," to appear in \textit{IEEE Communications Magazine}.
\end{textblock}

\bstctlcite{IEEEexample:BSTcontrol}
\title{How to Differentiate between Near Field and Far Field: Revisiting the Rayleigh Distance} 
\author{Shu Sun, Renwang Li, Chong Han, Xingchen Liu, Liuxun Xue, and Meixia Tao
%\thanks{Shu Sun, Renwang Li, Xingchen Liu, Liuxun Xue, and Meixia Tao are with the Department of Electronic Engineering and the Cooperative Medianet Innovation Center (CMIC), Shanghai Jiao Tong University, China. Chong Han is with Terahertz Wireless Communications (TWC) Laboratory, and with the Department of Electronic Engineering and CMIC, Shanghai Jiao Tong University, China. (\textit{Corresponding author: Chong Han})}
}

\maketitle

\begin{abstract}
Future wireless systems are likely to adopt extremely large aperture arrays to achieve higher throughput, wider coverage, and higher spatial resolution. Conventional wireless systems predominantly operate in the far field (FF) of the radiation source. However, as the array size increases and the carrier wavelength decreases, the near field (NF) becomes non-negligible. Since the NF and FF differ in many aspects, it is critical to identify their corresponding regions. In this article, we first provide a comprehensive overview of the existing NF-FF boundaries, then introduce a novel NF-FF demarcation method based on effective degrees of freedom (EDoF) of the channel. Since EDoF is intimately related to channel capacity, the EDoF-based border is able to characterize key channel performance more accurately than the classic Rayleigh distance and other representative benchmarks. Furthermore, we analyze the main features of the EDoF-based NF-FF boundary, provide insights into system design, and outline the associated challenges and research opportunities. 
\end{abstract}

\section{Introduction}
\IEEEPARstart{T}{he} deployment of fifth-generation (5G) networks on a global scale has prompted both academia and industry to concentrate on developing sixth-generation (6G) technologies. 6G is poised to meet the demands of emerging applications, such as integrated sensing and communications (ISAC), extended reality, holographic communications, and intelligent interaction, among others. To achieve the ambitious vision and requirements for 6G, numerous promising technologies have been proposed and are actively under investigation.

Notably, 6G is anticipated to leverage new spectrum resources, including the centimeter-wave, millimeter-wave, and terahertz bands. The higher propagation loss of these bands, as compared with the conventional sub-6 GHz spectrum, necessitates the deployment of antenna arrays to offer extra gain. Consequently, the number of antennas continues to increase, evolving from the standard 64 antennas to the scale of hundreds or even thousands. This technological advancement spawns the concept of extremely large aperture arrays (ELAAs), which offer the potential for high beamforming gain, exceptional spatial resolution, and remarkable spectral efficiency \cite{9170651}.

The deployment of ELAAs accentuates the near field (NF) effect. Generally, the electromagnetic (EM) field can be categorized into three regions: the \textit{reactive NF}, the \textit{radiative NF}, and the \textit{far field (FF)}. Since EM waves in the reactive NF are usually localized around the source and do not propagate, the radiative NF is more relevant to wireless communications, hence this article focuses on the radiative NF and refers to this as NF for simplicity. In the NF, EM wavefronts are spherical. Factors such as non-linear signal phase variations and differences in the traveling distance among different antennas in an ELAA need to be taken into account. The spherical wavefront model (SWM) considers the factors above and is thus highly accurate, but involves high computational complexity. Conversely, when the communication distance extends to the FF, EM wavefronts can be approximated by the planar wavefront model (PWM), since the angles at which the EM wave radiated from or received by each antenna in an array can be regarded as nearly identical. The PWM is a simplified model by neglecting the second- and higher-order of phase variations along an array, making it convenient for channel modeling, performance analysis, and system design. Therefore, the PWM has been widely adopted in the study of 5G and earlier networks. Nevertheless, due to the enormous array aperture of an ELAA, the NF extends significantly, expanding from just a few meters for a traditional array to hundreds of meters based on the classic Rayleigh distance (to be elaborated later on). Consequently, it is plausible for communication devices to situate in the NF. 

The NF and FF differ in many aspects, thus identifying their respective regimes has profound influence on wireless communications. \textcolor{black}{For instance, the primary impact is on channel capacity: The SWM in the NF can potentially enhance channel capacity due to increased rank of the channel, as compared with the PWM adopted in the FF~\cite{1510955}. As such, understanding these distinctions and accurately delineating the boundary between the NF and FF is crucial for optimizing capacity and ensuring efficient communication.} \textcolor{black}{In addition,} propagation channel properties such as path loss and spatial non-stationarity (SnS) can be quite distinct in the NF and FF. \textcolor{black}{Existing literature has unveiled that channel gain scales differently with distance in the NF and FF~\cite{Liu23OJCS}, it is thus critical to have accurate knowledge of the NF and FF regions so as to apply the corresponding power scaling laws.} Furthermore, SnS, referring to the variability of channel characteristics over space, is a prominent feature in the NF. It is usually caused by the large array aperture at the base station (BS), where different array elements observe varying scatterers or signal paths for the same user \textcolor{black}{such that signal amplitudes alter across the array elements~\cite{9170651,Chen23arXiv,elzanaty2023near}}. Accordingly, it is imperative to have an accurate evaluation of the NF-FF boundary to determine when the appropriate channel model should be adopted. \textcolor{black}{Moreover, recognizing the NF and FF is also relevant to sensing and localization~\cite{elzanaty2023near}.} Specifically, the antenna array response vector in the FF is solely influenced by the angle but not the distance, thus we can equally partition the angular domain to search for the best beam(s). The discrete Fourier-transform codebook can hence be employed to perform communication and sensing tasks, since each codeword in the codebook corresponds to a beam pointing to a certain angle. Besides, multiple anchors are often required to locate a target. While in the NF, the channel depends on both the angle and distance between the transmitter and receiver array elements, making the FF beamforming codebook unsuitable. On the other hand, the distance-dependent nature of spherical wavefronts facilitates the monostatic positioning of a target~\cite{Chen23arXiv}. Therefore, the NF-FF boundary is critical to selecting appropriate communication and sensing schemes in ISAC systems.

Although some NF-FF boundaries have been propounded in the literature, there lacks thorough review and comparison of various boundaries and their underlying criteria. Furthermore, only very limited investigation has been conducted on how to discern the NF and FF based on key performance indicators such as channel capacity. Towards this end, we first provide a comprehensive and in-depth overview of different categories of NF and FF boundaries, their demarcation metrics, and the impact on key communication performance measures (such as channel capacity). \textcolor{black}{Subsequently}, we propose a new NF-FF boundary based on effective degrees of freedom (EDoF) that is closely related to channel capacity \textcolor{black}{and yields expressions with each term holding a solid physical meaning}. We then analyze its behavior for multiple array configurations with diverse orientations and scattering environments, and compare it with representative benchmarks.

\section{Existing Research on Boundary between NF and FF}
As mentioned in the previous section, reasonable demarcation of the NF  and FF is momentous for wireless systems with ELAAs. In this section, we review existing research results on NF-FF boundaries, and classify them into four broad categories: phase-based, power-based, multiplexing-based, and capacity-based boundaries. The expressions for these NF-FF boundaries are provided in Table~\ref{tbl_dist} for ease of understanding and comparison. 

\subsection{Phase-Based Boundaries}
\subsubsection{Rayleigh Distance}
The classic border between the NF and FF of an antenna is called the \textit{Rayleigh distance} or \textit{Fraunhofer distance}, which is defined from the perspective of the phase error: If the distance between a transmitter and a receiver is larger than the Rayleigh distance, the maximum phase error across the antenna aperture between using the PWM and SWM is no more than $\pi/8$. This definition is easily extendable to antenna arrays, where the antenna aperture is substituted with the array aperture. The Rayleigh distance is commonly used to distinguish the NF and FF, since it can well capture the phase variations caused by the SWM and has a succinct mathematical form. However, the Rayleigh distance has weak association with channel performance metrics, such as channel gain and channel capacity that are important in practical wireless systems. 

\subsubsection{Advanced Rayleigh Distance}
For a multiple-input multiple-output (MIMO) system, regarding whether the NF MIMO channel can be modeled by the multiplication of NF array response vectors, the authors in \cite{Dai23TCOM} propose an MIMO advanced Rayleigh distance (MIMO-ARD), beyond which the largest phase discrepancy between the NF MIMO channel and the channel modeled by the multiplication of NF array response vectors is no more than $\pi/8$.

\begin{table*}[h!]
	\begin{center}
		\caption{Summary of NF-FF boundaries}\label{tbl_dist}
		\begin{threeparttable}
			\begin{tabular}{| c | c | c | c | c |}
				\hline 
				\textbf{Category} & \textbf{Boundary} & \textbf{\makecell{Array\\ configuration}} & \textbf{Criterion} & \textbf{Expression}  \\
				\hline \hline
				\multirow{2}{*}[-5ex]{\makecell{Phase\\-based}} & \makecell{Rayleigh distance} & \makecell{Arbitrary array\\ configurations} & \makecell*{Maximum phase error does not exceed $\pi/8$} & \makecell{Point to array: $\frac{2D^2}{\lambda}$\\Array to array: $\frac{2(D_\text{t}+D_\text{r})^2}{\lambda}$}  \\
				\cline{2-5}
				& \makecell{MIMO-ARD \cite{Dai23TCOM}} & \makecell{Arbitrary array\\ configurations} & \makecell*{Maximum phase error between the MIMO channel \\ and  the channel modeled by the multiplication of \\NF response vectors does not exceed $\pi/8$} & $\frac{4D_\text{t}D_\text{r}}{\lambda}$ \\
				\hline 
				
				\multirow{6}{*}[-2ex]{\makecell{Power\\-based}} 
				& \makecell{Critical \\ distance \cite{lu2021does}} & \makecell{Point to uniform\\ linear array (ULA)} & \makecell*{Power ratio between the weakest and strongest \\array elements remains above a specified threshold} &  $9D$  \\
				\cline{2-5}
				& \makecell{Uniform-power\\ distance \cite{9617121}} & \makecell{Point to uniform\\ planar array (UPA)} & \makecell*{Power ratio between the weakest and strongest \\array elements is no smaller than a threshold} & $\sqrt{\frac{\Gamma_\text{th}^{2/3}} {1- \Gamma_\text{th}^{2/3}}} \frac{D}{2}$  \\
				\cline{2-5}
				& \makecell{Effective Rayleigh \\ distance \cite{cui2021near}} & Point to ULA & \makecell*{Normalized beamforming gain under FF\\ assumption is no less than 95\%} & $0.367 \cos^2(\varphi) \frac{2D^2}{\lambda}$ \\
				\cline{2-5}
				& \makecell{Bj{\"o}rnson \\distance \cite{bjornson2021primer}} & \makecell{Point to UPA} & \makecell*{Normalized antenna array gain is close to 1} & $2L\sqrt{N}$  \\
				\cline{2-5}
				& \multirow{2}{*}[-0ex]{\makecell{Equi-power line\\ (surface) \cite{li2023applicable}}} & \makecell{Point to ULA} & \multirow{2}{*}[-0ex]{\makecell{Normalized received power is close to 1}} & \makecell{$2.86D$}  \\
				\cline{3-3} \cline{5-5}
				& & \makecell{Point to UPA} & & \makecell{$3.96D$}  \\
				\hline
			   \multirow{2}{*}[-2ex]{\makecell{Multi-\\plexing\\-based}}  
				
				& \makecell{Threshold\\ distance in \cite{4799060}} & ULA to ULA &  \makecell*{Ratio of the largest eigenvalues, as given by \\SWM and PWM, reaches a predefined threshold} & $\frac{\tau_\text{g} d_\text{t}d_\text{r}\cos(\vartheta_\text{t}) \cos(\vartheta_\text{r})}{\lambda}$  \\
				\cline{2-5} 
				& \makecell{Effective multiplexing\\ distance \cite{6800118}} & \makecell{ULA/uniform \\rectangular\\ array (URA) \\to ULA/URA} & \makecell*{Given an SNR, $m$ independent spatial streams \\can be accommodated simultaneously} & $\frac{D_\text{t}D_\text{r}} {\lambda \widetilde{S}^*_{\text{asy},m}(\gamma)}$ \\
				\hline 
				
				\multirow{1}{*}[-1ex]{\makecell{Capacity\\-based}} 
				
				& \makecell{Threshold distance \\in \cite{1510955}} & \makecell{ULA/URA \\to ULA/URA} & \makecell*{Capacity of the SWM surpasses that of the \\PWM by at least a factor of 1.5} &   {$\frac{\textcolor{black}{\epsilon} D_\text{t}D_\text{r}}{\lambda} \cos(\vartheta_\text{t}) \cos(\vartheta_\text{r})$} \\
				\hline \hline
				
				 \multirow{2}{*}[-0.5ex]{\makecell{EDoF\\-based}} 
				&  \multirow{2}{*}[-0.5ex]{\makecell{EBD \\ (Proposed)}} & \makecell{\textcolor{black}{ULA to ULA}}
				&  \multirow{2}{*}[-0.5ex]{\makecell*{\textcolor{black}{Ratio of EDoF $\Big(\frac{\text{tr}(\bf{R})}{\Vert\bf{R}\Vert_\text{F}}\Big)^2$ of SWM to PWM is} \\\textcolor{black}{larger than a predefined threshold}}}
				& \textcolor{black}{\makecell{${\frac{\pi D_\text{t}D_\text{r}}{\lambda a}} \vert\cos(\beta)-$ \\ $ \sin(\alpha)\sin(\alpha+\beta)\vert$}}\\ 
			    \cline{3-3} \cline{5-5}
			     & & \makecell{\textcolor{black}{URA to ULA}} & & \textcolor{black}{\makecell{{$\frac{\pi D_\text{r}\sqrt{D_{\text{t}_\text{x}}^2\text{sin}^2\left(\theta\right)+D_{\text{t}_\text{z}}^2\text{cos}^2\left(\theta\right)}}{\lambda b}$}}} \\
			    \hline
			\end{tabular}
			\begin{tablenotes}
				\footnotesize
				\item \noindent Note: $D$ denotes the array aperture, $\lambda$ denotes the wavelength, $\Gamma_\text{th}$ is a threshold smaller than one, $\varphi$ is the incident angle, $L$ represents the diagonal length of each antenna element, $N$ represents the number of antennas, $D_\text{t}$ and $D_\text{r}$ denote the array aperture at the transmitter and receiver, respectively, $\vartheta_\text{t}$ and $\vartheta_\text{r}$ denote the rotated angle at the transmitter and receiver, respectively, $d_\text{t}$ and $d_\text{r}$ denote the antenna spacing at the transmitter and receiver, respectively. $\tau_\text{g}$, $\widetilde{S}^*_{\text{asy},m}(\gamma)$, $\textcolor{black}{\epsilon}$, $a$, and $b$ are calculable auxiliary variables, where $a$ and $b$ are related to the numbers of antennas at the transmitter and receiver and the EDoF ratio threshold, $\bf{R}$ denotes the covariance matrix of MIMO channel matrix, $\text{tr}(\cdot)$ and $\Vert\cdot\Vert_\text{F}$ denote the trace and Frobenius norm of a matrix, respectively, $D_{\text{t}_\text{x}}$ and $D_{\text{t}_\text{z}}$ respectively denote the horizontal and vertical lengths of the URA. \textcolor{black}{The remaining notations are aligned with those in Fig.~\ref{fig_arrays}.}
			\end{tablenotes}
		\end{threeparttable}
	\end{center}
\end{table*}

 \subsection{Power-Based Boundaries}
 \subsubsection{Critical Distance}
 The Rayleigh distance primarily pertains to the maximum acceptable phase difference among array elements. However, when utilizing maximum ratio combining (MRC), signal phases can be perfectly aligned, eliminating their impact on the received power. In this context, the received power relies purely on the amplitude response. Consequently, the authors in \cite{lu2021does} propound a critical distance beyond which the power ratio between the weakest and strongest array elements remains above a certain threshold.

 \subsubsection{Uniform-Power Distance}
 The critical distance in \cite{lu2021does} is extended to the uniform-power distance (UPD) in \cite{9617121} by additional consideration of the uniform planar array (UPA) structure and the variation of the projected aperture across the array. The UPD maintains the same definition as in \cite{lu2021does}, defining a distance beyond which the power ratio between the weakest and strongest array elements is no smaller than a threshold. 
 
 \subsubsection{Effective Rayleigh Distance}
 The authors in \cite{lu2021does} and \cite{9617121} adopt MRC to eliminate the influence of signal phases. However, due to the inherent challenges in achieving perfect channel estimation, the MRC may not completely cancel out the phases. Therefore, from a beamforming perspective, the authors in \cite{cui2021near} propose an effective Rayleigh distance, beyond which the normalized beamforming gain under the FF assumption is no less than 95\%. The effective Rayleigh distance can be regarded as a correction for the Rayleigh distance to ensure the beamforming gain when adopting the PWM.
 
\subsubsection{Bj{\"o}rnson Distance}
The authors in \cite{bjornson2021primer} consider a UPA consisting of identical antennas each with an equal area. From a strict EM perspective, they introduce a normalized array gain, representing the received power relative to the power obtained using the PWM. Under this modeling, the Rayleigh distance can be reformulated considering the diagonal length of each array element. Then, the Bj{\"o}rnson distance is proposed, beyond which the maximum array gain can be achieved. 

\subsubsection{Equi-Power Line/Surface}
Considering the uniform linear array (ULA) architecture with a line-of-sight (LoS) path, the received power under the MRC from a point source is solely determined by the amplitude response and is independent of the signal phase.  The authors in \cite{li2023applicable} thereby define a ratio of the received power obtained with the SWM to that obtained with the PWM. When the communication distance grows to infinity, the ratio approaches one. They propound an equi-power line where this ratio reaches a predefined threshold. An equi-power surface is defined similarly for the UPA. 

\subsection{Multiplexing-Based Boundaries}
\subsubsection{Threshold Distance in \cite{4799060}}
Building upon an approximation of the largest eigenvalue of LoS MIMO channels employing the ULA, the authors in \cite{4799060} obtain a threshold distance at which the ratio of the largest eigenvalues given by the SWM and PWM, respectively, reaches a predefined threshold. This threshold distance is proportional to the product of the array apertures at the transmitter and receiver, inversely proportional to the wavelength, and also has to do with the orientations of the arrays. 

\subsubsection{Effective Multiplexing Distance}
Considering spatial multiplexing, the authors in \cite{6800118} propose an effective multiplexing distance (EMD), which represents the farthest range at which the channel can efficiently accommodate $m$ independent spatial streams simultaneously at a specific signal-to-noise ratio (SNR). A key parameter influencing the EMD is the ratio of the $m$-th largest eigenvalue of the channel covariance matrix to the first largest eigenvalue. 

 \subsection{Capacity-Based Boundaries}
% \subsubsection{Threshold Distance in \cite{1510955}}
 The authors in \cite{1510955} propose a threshold distance below which the capacity of the SWM surpasses that of the PWM by at least a factor of 1.5, for a given array size. In practical terms, this threshold distance marks a point where the capacity underestimation error when using the PWM is 50\%. This threshold distance can also be regarded as the 0.225 dB-down beamwidth distance of the array. 
 
 \subsection{Summary}
One of the most relevant performance metrics for MIMO systems is channel capacity. However, the aforementioned NF-FF boundaries either have no linkage to channel capacity, or the linkage is too implicit to quantify. In the next section, we introduce a new NF-FF demarcation criterion leveraging the EDoF of the channel, which is closely related to both the multiplexing capability and channel capacity. Although the threshold distance in \cite{1510955} also considers the channel capacity estimation error, the boundary is obtained empirically and the error tolerance is relatively large. While the EMD in \cite{6800118} involves DoF as well, it regards a sub-channel as a valid one if its eigenvalue exceeds a certain threshold. In contrast, the EDoF can be approximately interpreted as the effective number of sub-channels with equal power, hence is more likely to be higher than one as compared with that in \cite{6800118}, thus rendering larger boundary distances and more accurate capacity estimation as will be evident in the simulation results later. \textcolor{black}{Note that concepts and types of EDoF have been presented in recent work relevant to NF communications, such as~\cite{Ouyang23near}} and references therein, offering insightful guidance on NF system design and analysis. This article focuses more on how to leverage the EDoF to identify NF and FF regimes. As mentioned in~\cite{Ouyang23near}, different types of EDoF embody discrepant physical meanings and application scopes, the EDoF adopted herein is closely related to channel capacity while facilitating quantitative evaluation of the NF-FF boundary. Furthermore, we provide succinct expressions of the EDoF-based boundary distance (EBD) for various array configurations, \textcolor{black}{including ULA and uniform rectangular array (URA) architectures.}

\section{Proposed Demarcation Based on Effective Degrees of Freedom}
\subsection{Demarcation Criterion}
%As mentioned above, channel capacity is closely related to the EDoF of the channel, where the EDoF represents the equivalent number of single-input-single-output sub-channels. 

In order to characterize channel performance (e.g., channel capacity) more accurately, we introduce a novel NF-FF demarcation approach based upon the EDoF, whose information-theory-originated definition \cite{Muharemovic08TIT} is shown in the last row of Table~\ref{tbl_dist}. Specifically, the boundary between the NF and FF is defined such that the EDoF ratio of the SWM to the PWM of an MIMO system equals a predefined threshold at the boundary. The threshold value of the EDoF ratio may depend on the tolerance for capacity estimation errors and can usually be set as slightly larger than one. A smaller threshold implies closer proximity between the SWM and PWM. In this context, we set the threshold to 1.01 and 1.05 in the simulations later on, to obtain the upper and lower bounds of the EBD, respectively, demonstrating representative outcomes compared to alternative boundaries. This EDoF demarcation criterion can capture the differential spatial characteristics of MIMO systems between the SWM and PWM, and is more explicitly related to key performance indicators such as the multiplexing order and channel capacity~\cite{Muharemovic08TIT}. 

\subsection{Case Studies}\label{sec_EDoF_caseStudy}
\begin{figure}[t!]
	\begin{centering}
		\includegraphics[width=0.44\textwidth]{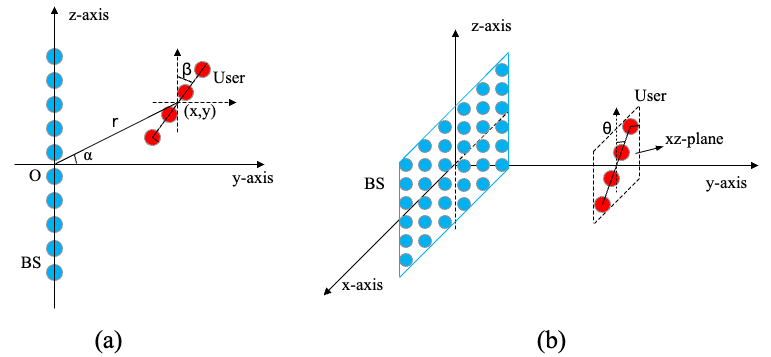}
		\caption{\textcolor{black}{Array configurations. (a) Both the BS and user are equipped with a ULA, where $\alpha$ denotes the angle between the centers of the BS and user ULAs, and $\beta$ is the angle of the user ULA against the positive Z-direction within the YZ-plane. (b) The BS and user are equipped with a URA and a ULA, respectively, where $\theta$ represents the angle of the ULA against the positive Z-direction and parallel with the XZ-plane. }}\label{fig_arrays}
	\end{centering}
\end{figure}

In the sequel, we will unveil the main behaviors of the proposed EBD and how it differs from the classic Rayleigh distance and other representative boundaries through some examples. \textcolor{black}{We consider point-to-point MIMO systems, as illustrated in Fig.~\ref{fig_arrays}, where the user is equipped with a ULA, and the antenna array at the BS can be a ULA or URA.}

\textcolor{black}{Based upon the aforementioned EDoF, we conduct mathematical derivations for the two types of MIMO systems in Fig.~\ref{fig_arrays} to obtain expressions of the EBD, and provide the results in the last row of Table~\ref{tbl_dist}.} Note that besides the array apertures and wavelength, the EBD is also related to the number of transmit and receive antennas. The quantitative influence of the number of antennas on the EBD is complicated, but in general the EBD first decreases rapidly and then saturates gradually as the number of antennas increases given a fixed array aperture, thus the boundary for an array with two antennas serves as an upper bound.

\begin{figure}[t!]
	\begin{centering}
		\includegraphics[width=0.4\textwidth]{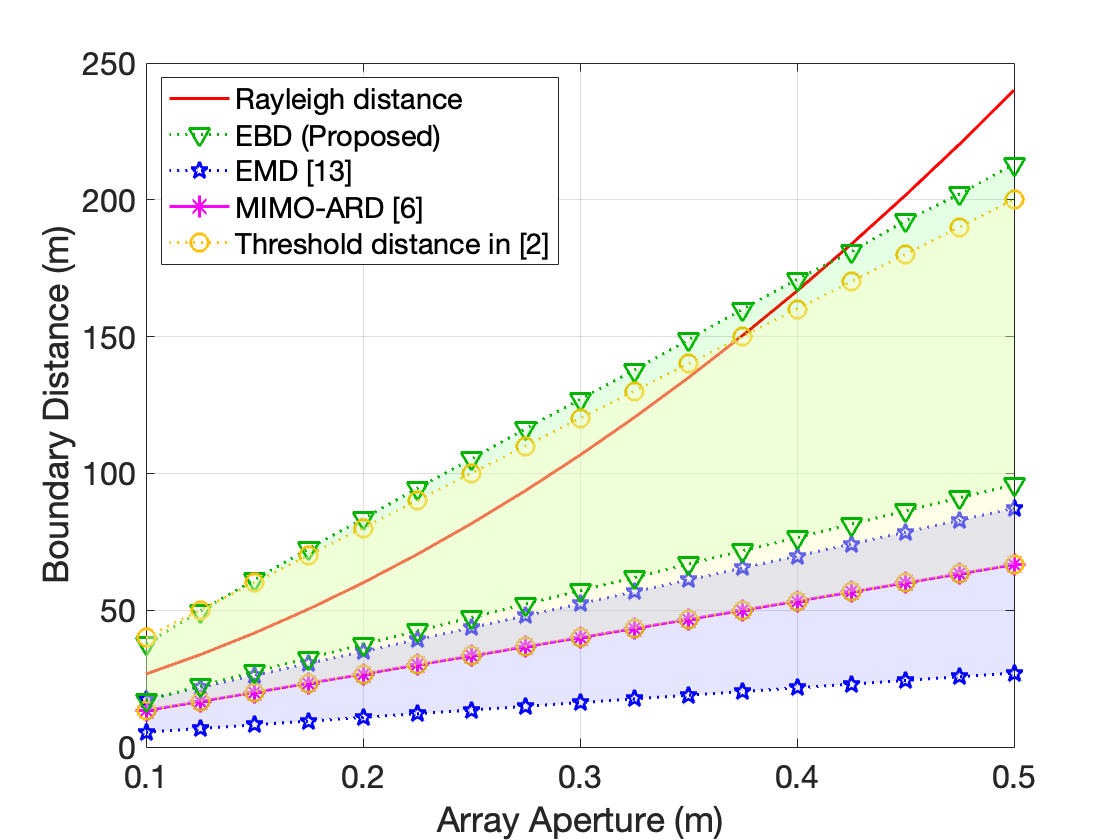}
		\caption{Comparison of different NF-FF boundaries, where the receiver is equipped with a two-element ULA with a length of 0.1 m, the transmitter has a URA with half-wavelength element spacing and a horizontal length of 0.05 m, and the abscissa values denote the maximum aperture of the URA. The orientation angle $\theta$ in Fig.~\ref{fig_arrays}(b) is 0$^\circ$. Shaded areas correspond to the boundary criteria with varying threshold values. Specifically, to obtain the upper and lower bounds of the corresponding boundaries, the EDoF threshold is respectively set to 1.01 and 1.05 for the EBD; the ratio of the second largest eigenvalue to the first one is respectively set to -20 dB and -10 dB for the EMD in \cite{6800118}; \textcolor{black}{the threshold factor is respectively set to 4 and 12 for the threshold distance in~\cite{1510955}.}}\label{fig_dist_aperture}
	\end{centering}
\end{figure}

\begin{figure}[t!]
	\begin{centering}
		\includegraphics[width=0.4\textwidth]{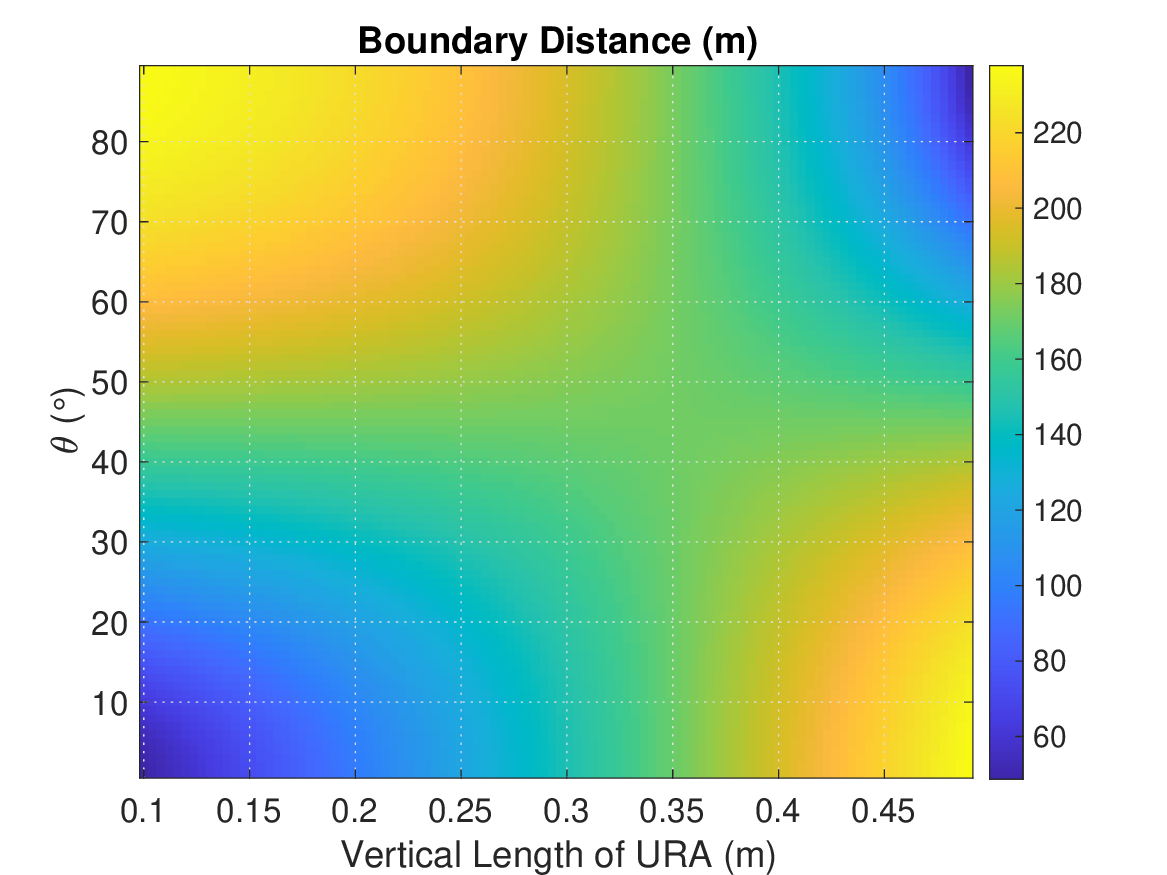}
		\caption{Variation of the EBD with the orientation angle $\theta$ and the vertical length of the URA depicted in Fig.~\ref{fig_arrays}(b). The URA aperture is 0.5 m, and the numbers of elements along the horizontal and vertical directions of the URA are both eight. The ULA has two elements and its length is 0.1 m. The EDoF threshold between the SWM and PWM is set to 1.01.}\label{fig_dist_angle}
	\end{centering}
\end{figure}

Next, we provide numerical analysis on the EBD to gain more intuitive insights. In all the simulations, the carrier frequency is set to 100 GHz. Fig.~\ref{fig_dist_aperture} compares various NF-FF boundaries. We consider the URA-to-ULA case which is common in the practical BS-to-user communication scenario, and select the baselines aligned with such a scenario. Several relevant observations can be drawn: First, the proposed EBD, along with some other boundaries, can vary within a certain range depending on the threshold of the key parameter involved. For the EBD, its EDoF ratio threshold should be reduced as the tolerance for the capacity estimation error decreases. Regardless of the threshold, all the boundaries studied grow approximately linearly with the array aperture, except for the Rayleigh distance, and the threshold mainly determines the slope. Second, the Rayleigh distance can be smaller or larger than the EBD, implying that it may underestimate or overestimate the NF region from the viewpoint of supportable spatial streams. Moreover, the EBD is larger than the other benchmarks in most cases. Since the PWM becomes more accurate as the link distance increases, the channel capacity calculated using the PWM is also closer to that computed using the SWM. Therefore, the EBD leads to a tighter approximation to the true channel capacity. Fig.~\ref{fig_dist_angle} illustrates the variation of the EBD with the orientation angle $\theta$ and the vertical length of the URA depicted in Fig.~\ref{fig_arrays}(b). It is observed that given a fixed aperture, the EBD increases with $\theta$ when the vertical length of the URA is small, and behaves oppositely for a large vertical length. This is because when the vertical length is small, the $\text{sin}^2\left(\theta\right)$ term in the EBD expression in Table~\ref{tbl_dist} dominates, hence the EBD increases with $\theta$ for $\theta\in\left(0^\circ,90^\circ\right)$, and vice versa. Additionally, there exists a rotation center at around ($0.35$ m, $45^\circ$), corresponding to equal horizontal and vertical lengths of the URA.

\begin{figure}[t!]
	\begin{centering}
		\includegraphics[width=0.4\textwidth]{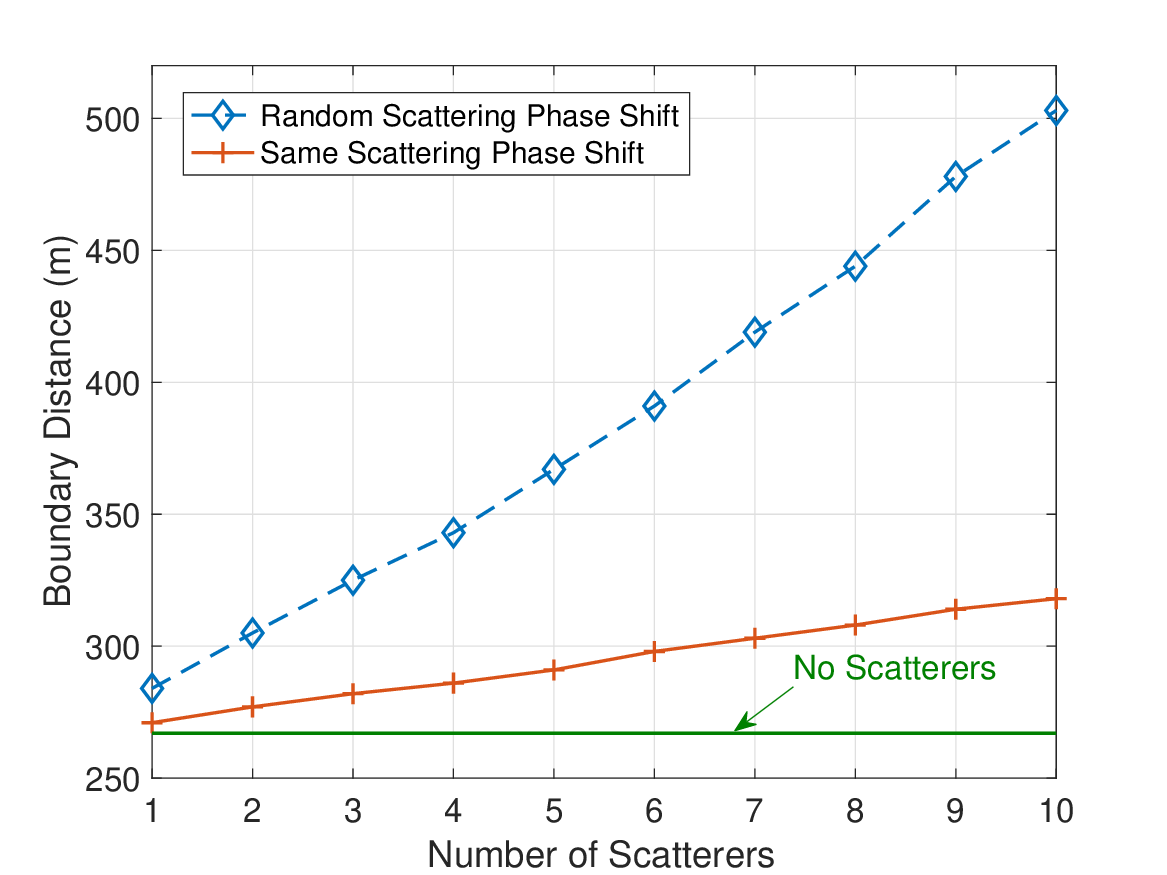}
		\caption{Variation of the EBD with the number of scatterers for two ULAs. The ULAs are broadside to each other, with lengths of 2 m and 0.1 m, and numbers of elements of 16 and 8, respectively. The EDoF threshold between the SWM and PWM is set to 1.05.}\label{fig_dist_sca}
	\end{centering}
\end{figure}

In practice, scatterers usually exist between the BS and user, creating multipath environments, it is thus worth exploring how the NF-FF boundary changes as opposed to the pure-LoS-path scenario. Variations of the EBD with the number of scatterers are displayed in Fig. \ref{fig_dist_sca}, where the amplitude of each scattered path is proportional to the inverse of the product of the distances between the transmitter and scatterer and between the scatterer and \textcolor{black}{receiver.} Since scattering may cause phase shift and attenuation to incident EM waves, and the amount of phase shift and attenuation depends on scatterer properties, we study two extreme cases: (1) random scattering phase shift, where phase shift for the path between each transmit and receive antenna pair is independently and randomly generated, and (2) identical phase shift for the scattered paths corresponding to the same incident path. The scattering attenuation is between 0 and 1. Fig.~\ref{fig_dist_sca} reveals that the EBD increases with the number of scatterers and with the randomness of scattering phase shifts, assuming each scatterer is unobstructed from every antenna at the transmitter and receiver. This is because more significant phase variation leads to higher degrees of non-linearity of phase shift across the array elements, it hence takes a longer distance to approximate the linear phase shift in the PWM, indicating that the EBD tends to increase with the diversity of scatterer types. Nevertheless, when the power ratio of LoS and scattered paths approaches a constant level as in many practical systems, the EBD remains almost unaltered with the number of scatterers.

\begin{figure} 
	\centering
	\subfloat[\label{fig:a}]{
		\includegraphics[scale=0.4]{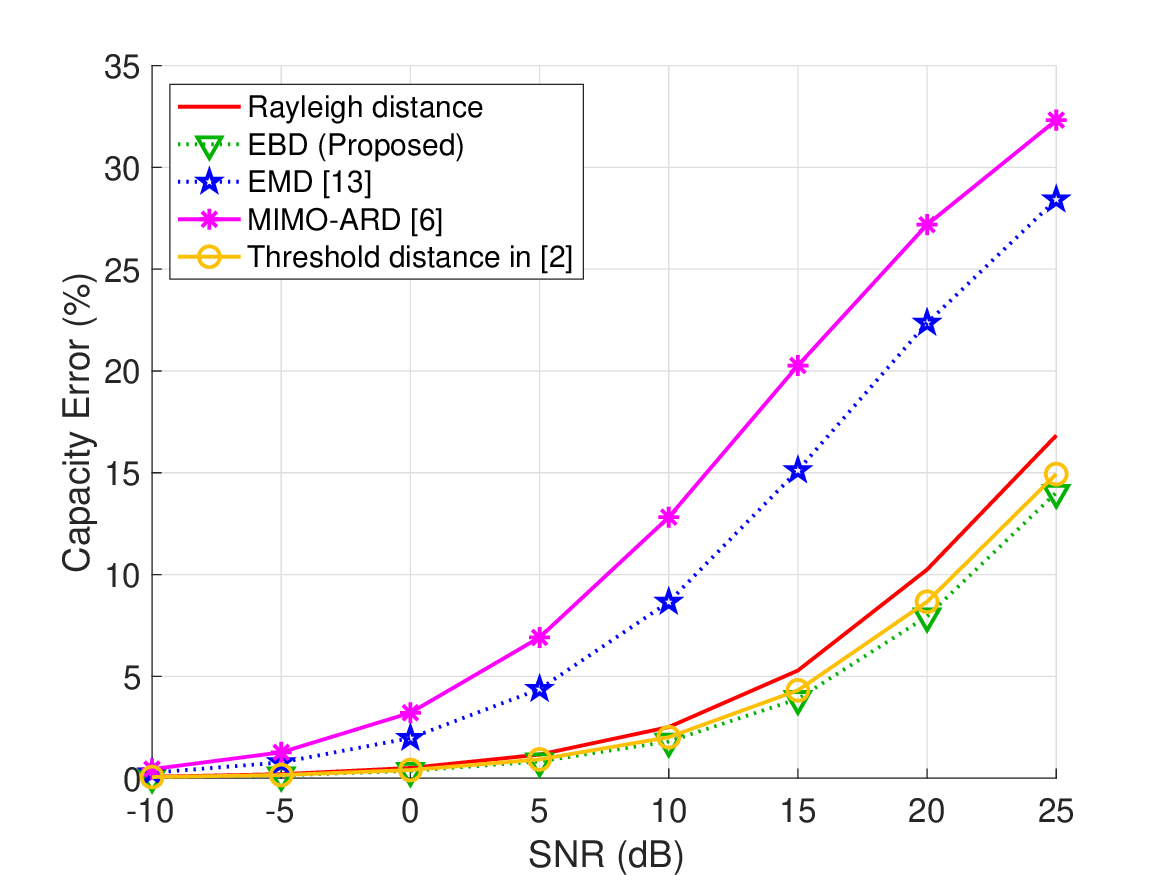}}\\
	\subfloat[\label{fig:b}]{
		\includegraphics[scale=0.4]{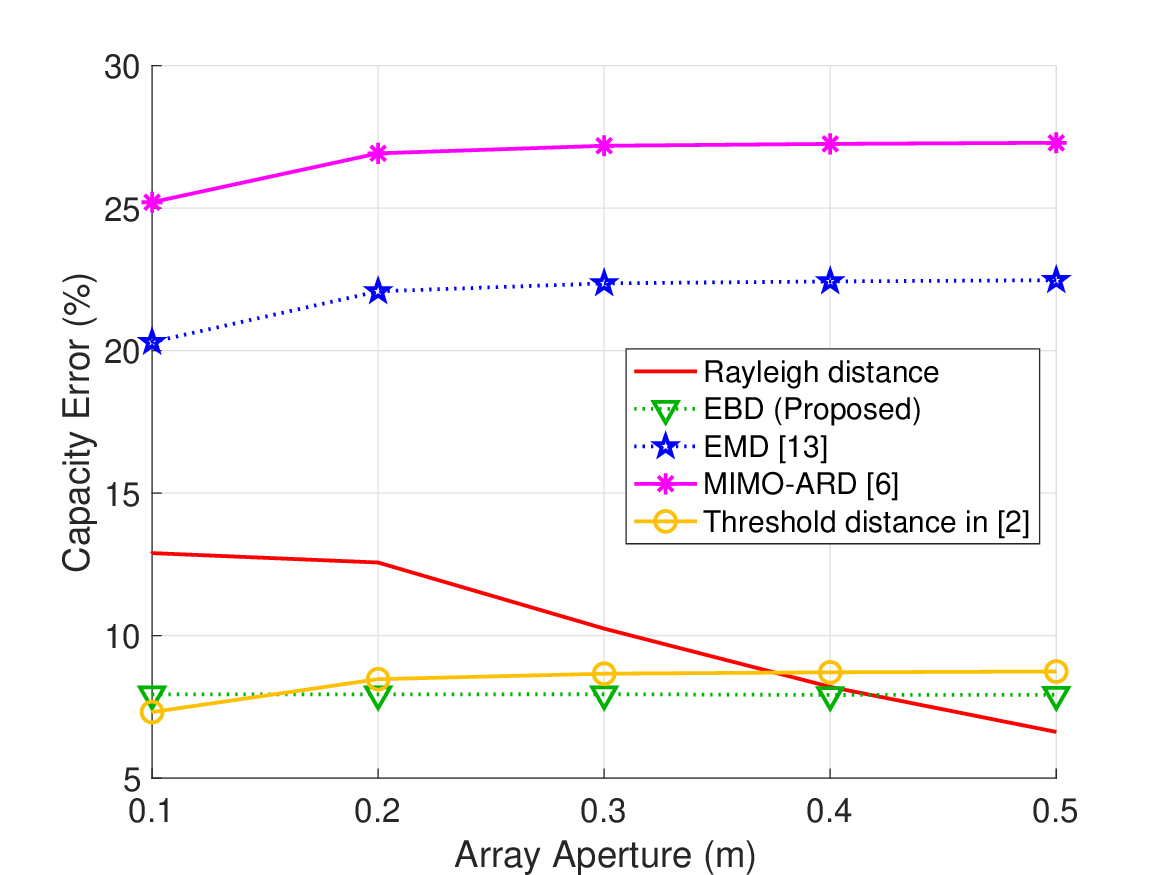}} 
	\caption{\textcolor{black}{Channel capacity estimation error for various NF-FF boundaries for the ULA-to-URA scenario against (a) SNR and (b) URA aperture. The ULA length is 0.1 m. The URA aperture is fixed to 0.3 m in (a), while the SNR is 20 dB in (b). For boundary criteria with varying threshold values, those corresponding to the upper bounds in Fig.~\ref{fig_dist_aperture} are adopted.}}
	\label{fig_cap} 
\end{figure}

\section{Implications for System Design and Research Opportunities}
In this section, we elaborate on the implications of the NF-FF boundary on relevant aspects of wireless systems, and highlight potential directions for future research.
\subsection{Implications for System Design}
\subsubsection{Channel Capacity}
%The NF-FF boundary has a crucial impact on channel capacity, since it determines the applicable regions of different EM wavefronts which in turn dictate the channel model. 
It has been shown in Fig.~\ref{fig_dist_aperture} that the classic Rayleigh distance may underestimate or overestimate the NF zone compared with the EBD, whereas other boundaries often underestimate. Overestimation of the NF range can induce unnecessary computational complexity using the SWM, while underestimation may cause prediction error of channel capacity. For instance, Fig.~\ref{fig_cap} quantifies the estimation error of channel capacity for the URA-to-ULA scenario, where the estimation error is computed by comparing the channel capacity using the PWM with that using the SWM at the corresponding boundary distance. As evident from Fig.~\ref{fig_cap}, some NF-FF boundaries cause large estimation error in channel capacity and the error grows with the SNR. \textcolor{black}{Although the threshold distance in~\cite{1510955} yields comparable errors as the EBD, its threshold factor in this case is drastically larger than the typical value and lacks solid physical meaning.} Moreover, the estimation error of the other boundaries varies with the array aperture, especially the Rayleigh distance, while the error alters little against the entire aperture range considered for the EBD. Therefore, there can be significant differences in both the estimation error and how the error changes with system configurations, when using different NF-FF demarcation methods, thus the selection of the boundary is doubly crucial for performance evaluation. 

\subsubsection{Multiple Access} 
To gain insights into how the NF-FF boundary influences multiple access, we regard the two-element ULA in the aforementioned simulations as two users each with a single antenna. In this sense, the boundary distance in Fig.~\ref{fig_dist_angle} implies the distance at which the spatial channels between the two users are approximately fully correlated. It can be inferred from Fig.~\ref{fig_dist_angle} that the spatial channels between two users are less correlated when the line connecting them is parallel to the longer side of a URA. Consequently, if solely aiming to serve more users in the NF, it is beneficial to place the URA at the BS such that its longer side is parallel to the direction with most users. In reality, other factors also need to be considered in system design. Thus, knowledge of how the NF-FF boundary varies with the azimuth and elevation angles is helpful in determining array orientations and designing adaptive algorithms for channel estimation and beam management, so as to serve users more efficiently based on their locations and relative orientations. 

\subsection{Challenges and Opportunities}
For the EBD, it is worth continuing to explore succinct mathematical expressions of the boundary considering more categories of array configurations and rotation angles. Although numerical simulations can be employed to obtain accurate boundary results, it is usually time-consuming and can barely provide intuitive guidance on system design and evaluation. Furthermore, the EDoF involved in this work is dictated by the channel matrix structure relying solely on the link distance, hence is applicable to arbitrary SNRs. Nevertheless, the approximation error of the EDoF-based capacity is non-negligible at high SNRs. This may be compensated for by setting a smaller EDoF threshold (i.e., closer to 1) when determining the EBD, but other approaches are also worth pursuing, such as studying new demarcation methodologies capable of more accurately characterizing channel capacity and other performance indicators while remaining mathematically analyzable.

%Moreover, it is seen from Fig.~\ref{fig_cap} that although the EBD yields more accurate capacity estimation as compared with the benchmarks, the estimation error is still considerable at high SNRs. Therefore, it is valuable to further improve the EBD or to study new demarcation methodologies that are capable of more accurately characterizing channel capacity and other performance indicators while remaining mathematically analyzable. 

Additionally, there are various definitions of EDoF~\cite{Ouyang23near}. When adopting other types of EDoF for acquiring NF-FF boundaries, the exploitable EDoF may increase with the SNR, and can be identical to or different from its SNR-independent counterpart~\cite{Ouyang23near}. Qualitatively speaking, the NF-FF boundary distance may decrease as compared to its SNR-independent counterpart for low SNRs, and vice versa. Quantitatively, the NF-FF boundary distance can be obtained by judiciously setting the EDoF threshold according to the operating SNR, which deserves further investigation.

\section{Conclusion}
In this article, we discussed the importance of identifying the NF regime for ELAAs, summarized existing NF-FF boundaries, and propounded a novel NF-FF demarcation scheme based on the EDoF of the MIMO channel. We investigated key influencing factors and behaviors of the EBD for various array configurations and scattering environments, and analyzed the implications for system design. The proposed NF-FF boundary is able to more accurately characterize system performance indicators such as channel capacity, as compared with representative baselines, and can provide more insights into system design. However, there are still some challenges to tackle and research opportunities to exploit as discussed above.

\bibliographystyle{IEEEtran}
\bibliography{review}

% Generated by IEEEtran.bst, version: 1.14 (2015/08/26)
\begin{thebibliography}{10}
\providecommand{\url}[1]{#1}
\csname url@samestyle\endcsname
\providecommand{\newblock}{\relax}
\providecommand{\bibinfo}[2]{#2}
\providecommand{\BIBentrySTDinterwordspacing}{\spaceskip=0pt\relax}
\providecommand{\BIBentryALTinterwordstretchfactor}{4}
\providecommand{\BIBentryALTinterwordspacing}{\spaceskip=\fontdimen2\font plus
\BIBentryALTinterwordstretchfactor\fontdimen3\font minus
  \fontdimen4\font\relax}
\providecommand{\BIBforeignlanguage}[2]{{%
\expandafter\ifx\csname l@#1\endcsname\relax
\typeout{** WARNING: IEEEtran.bst: No hyphenation pattern has been}%
\typeout{** loaded for the language `#1'. Using the pattern for}%
\typeout{** the default language instead.}%
\else
\language=\csname l@#1\endcsname
\fi
#2}}
\providecommand{\BIBdecl}{\relax}
\BIBdecl

\bibitem{9170651}
E.~D. Carvalho \emph{et~al.}, ``Non-stationarities in extra-large-scale massive
  {MIMO},'' \emph{IEEE Wireless Communications}, vol.~27, no.~4, pp. 74--80,
  Aug. 2020.

\bibitem{1510955}
J.-S. Jiang and M.~Ingram, ``Spherical-wave model for short-range {MIMO},''
  \emph{IEEE Transactions on Communications}, vol.~53, no.~9, pp. 1534--1541,
  Sep. 2005.

\bibitem{Liu23OJCS}
Y.~Liu \emph{et~al.}, ``Near-field communications: A tutorial review,''
  \emph{IEEE Open Journal of the Communications Society}, vol.~4, pp.
  1999--2049, 2023.

\bibitem{Chen23arXiv}
H.~Chen \emph{et~al.}, ``6{G} localization and sensing in the near field:
  Features, opportunities, and challenges,'' \emph{arXiv preprint
  arXiv:2308.15799}, Accessed on 3 Apr. 2024.

\bibitem{elzanaty2023near}
A.~Elzanaty \emph{et~al.}, ``Near and far field model mismatch: Implications on
  {6G} communications, localization, and sensing,'' \emph{arXiv preprint
  arXiv:2310.06604v2}, Accessed on 8 Jul. 2024.

\bibitem{Dai23TCOM}
Y.~Lu and L.~Dai, ``Near-field channel estimation in mixed {LoS/NLoS}
  environments for extremely large-scale {MIMO} systems,'' \emph{IEEE
  Transactions on Communications}, vol.~71, no.~6, pp. 3694--3707, Jun. 2023.

\bibitem{lu2021does}
H.~Lu and Y.~Zeng, ``How does performance scale with antenna number for
  extremely large-scale {MIMO}?'' in \emph{2021 IEEE International Conference
  on Communications}, Montreal, QC, Canada, 2021, pp. 1--6.

\bibitem{9617121}
H.~Lu and Y.~Zeng, ``Communicating with extremely large-scale array/surface:
  Unified modeling and performance analysis,'' \emph{IEEE Transactions on
  Wireless Communications}, vol.~21, no.~6, pp. 4039--4053, Jun. 2022.

\bibitem{cui2021near}
M.~Cui and L.~Dai, ``Near-field wideband beamforming for extremely large
  antenna arrays,'' \emph{IEEE Transactions on Wireless Communications (Early
  Access)}, pp. 1--1, 2024.

\bibitem{bjornson2021primer}
E.~Bj{\"o}rnson \emph{et~al.}, ``A primer on near-field beamforming for arrays
  and reconfigurable intelligent surfaces,'' in \emph{2021 55th Asilomar
  Conference on Signals, Systems, and Computers}, Pacific Grove, CA, USA, 2021,
  pp. 105--112.

\bibitem{li2023applicable}
R.~Li \emph{et~al.}, ``Applicable regions of spherical and plane wave models
  for extremely large-scale array communications,'' accepted to \textit{China
  Communications}. arXiv preprint arXiv:2301.06036. Accessed on 3 Apr. 2024.

\bibitem{4799060}
F.~Bohagen \emph{et~al.}, ``On spherical vs. plane wave modeling of
  line-of-sight {MIMO} channels,'' \emph{IEEE Transactions on Communications},
  vol.~57, no.~3, pp. 841--849, Mar. 2009.

\bibitem{6800118}
P.~Wang \emph{et~al.}, ``Tens of gigabits wireless communications over {E-Band}
  {LoS} {MIMO} channels with uniform linear antenna arrays,'' \emph{IEEE
  Transactions on Wireless Communications}, vol.~13, no.~7, pp. 3791--3805,
  Jul. 2014.

\bibitem{Ouyang23near}
C.~Ouyang \emph{et~al.}, ``Near-field communications: A degree-of-freedom
  perspective,'' \emph{arXiv preprint arXiv:2308.00362}, Accessed on 3 Apr.
  2024.

\bibitem{Muharemovic08TIT}
T.~Muharemovic \emph{et~al.}, ``Antenna packing in low-power systems:
  Communication limits and array design,'' \emph{IEEE Transactions on
  Information Theory}, vol.~54, no.~1, pp. 429--440, Jan. 2008.

\end{thebibliography}

\end{document}